\documentclass[fleqn,12pt,twoside]{article}
\usepackage{espcrc1}

\usepackage{epsfig}
\usepackage{graphicx}
\usepackage[figuresright]{rotating}

\newcommand{\be}{\begin{equation}}
\newcommand{\ee}{\end{equation}}
\newcommand{\ba}{\begin{eqnarray}}
\newcommand{\ea}{\end{eqnarray}}

\renewcommand{\>}{\rangle}

\newcommand{\AmS}{{\protect\the\textfont2
  A\kern-.1667em\lower.5ex\hbox{M}\kern-.125emS}}
\def\spose#1{\hbox to 0pt{#1\hss}}
\def\ltapprox{\mathrel{\spose{\lower 3pt\hbox{$\mathchar"218$}}
 \raise 2.0pt\hbox{$\mathchar"13C$}}}

\def\PR{{ Phys.\ Rev.\ }}

\def\lsim{\raise0.3ex\hbox{$<$\kern-0.75em\raise-1.1ex\hbox{$\sim$}}}
\def\gsim{\raise0.3ex\hbox{$>$\kern-0.75em\raise-1.1ex\hbox{$\sim$}}}

\title{ A new order parameter with renormalized Polyakov loops
      \thanks{This work was supported by DFG Grant No. FOR 339/1-2.} }

\author{F. Zantow with O. Kaczmarek, F. Karsch, P. Petreczky\\\vspace{0.0cm}
       {Fakult\"at f\"ur Physik, Universit\"at Bielefeld,
        D-33615 Bielefeld, Germany}}

\begin{document}
\maketitle
\vspace{-0.3cm}
\section{INTRODUCTION}
\vspace{-0.2cm}
It is believed that hadronic matter undergoes a confinement/deconfinement phase
transition at sufficiently high temperatures $T$ and/or densities $n_B$. In lattice calculations the
expectation value of the Polyakov loop $L$ is treated as an order parameter for
this phase transition. Physically, however, this quantity has no meaning since
$T\ln|\<L\>|$ refers to the free energy of an isolated heavy quark and contains
divergent self-energy contributions. Instead of $T\ln|\<L\>|$ a meaningful determination of the free
energy of a heavy quark pair is proposed in [1]. We use this
determination to construct
a physical order parameter for the mentioned phase transition.
As a first test of our investigation we refer to lattice studies of the pure $SU(3)$
 theory [1].
\vspace{-0.2cm}
\section{COLOR AVERAGING AT DIFFERENT ENERGY SCALES}
\label{section:results}
\vspace{-0.2cm}
 We have calculated the color averaged free energy $F_{av}$ of
a static heavy quark pair using Polyakov loop correlations
[2] and the color singlet free energy
$F_s$ using the time-like closed
Wilson loops [3] at finite temperature.
The color averaged free energy is related to the color singlet ($F_s$) and octet
($F_o$) contributions: $\mbox{exp}(-F_{av}/T)=1/9\mbox{exp}(-F_s/T)+8/9\mbox{exp}(-F_o/T)$.
In a static system the internal energy reduces to the potential
$V$ and therefore the free energy $F$ refers to the potential, $F=V-TS$,
where we assume that the entropy $S$ may stay $R$-dependent ($R$: distance).
In a perturbative treatment of QCD the singlet is of attractive
while the octet potential is of repulsive Coulomb form [4]. In the
hadronic phase additional linear terms contribute to the potentials and signal
confinement.

 For $RT<<1$, the repulsive octet contribution to the color averaged free energy
is exponentially suppressed and the linear terms become negligible; hence we find
\begin{eqnarray}
\lim_{RT \to 0}\left\{F_{av}(R,T)-F_s(R,T)\right\}&=&T\ln 9.
\end{eqnarray}
 Due to this relation both free energies coincide at $T=0$. Moreover (1) predicts a
 singlet-like behavior of $F_{av}$ at short distances [1]. We argue
that the difference $T\ln 9$ is caused by the
difference of the entropy contributions; $\ln 9$ has the typical
form of entropies if one identifies $9$ with the sum over the nine possible
 color configurations: $S_{s}-S_{av}=\ln 9$.

\indent For $RT>>1$ and $T>T_c$, the contributions from the potentials become
negligible and the color averaging leads to $S_{av}=\ln\left\{1/9\;\mbox{exp}(S_s)+8/9\;\mbox{exp}(S_o)\right\}.$
For $T<T_c$, however, the contributions from the potentials and entropies to $F_{av}$ cannot
easily be separated because of the linear confining terms in the potentials.

\indent We have studied (1) with our lattice data [1] in terms of the
difference $(F_{av}-F_s)/T$ in Fig.~1a. The data approach the
value $\ln 9$ in both phases at distances $RT\lsim 0.1$. At these distances the
averaged potential behaves like a singlet one. At large $RT$, however, the difference vanishes although $F_{av}$ and $F_s$
separately approach non-zero constants for $T>T_c$ and diverge for $T<T_c$.
We note that the deviations of $(F_{av}-F_s)/T$ from $\ln 9$ at
large RT indicate that octet contributions to the averaged free energy
are important at all temperatures. In fact, Fig.~1a and the above mentioned
considerations suggest $S_{av}=S_{s}=S_o$ for large $R$ and $T>T_c$.
\vspace{-0.2cm}
\section{THE RENORMALIZED POLYAKOV LOOP}
\label{section:results}
\vspace{-0.2cm}
In [1] the renormalized free energy of a heavy quark pair is obtained by matching
the lattice data to the $T=0$ potential at short distances. With respect to (1)
$F_{av}$ and $F_s$ coincide at large distances and yield the free energy of an infinitely separated
quark pair (Fig.~1a).
We suggest that $L^{ren}\equiv\mbox{exp}(-F_{av}(\infty)/2T)$, where $F_{av}(\infty)$ has to
be taken from normalized data, can be used as an order
parameter for the deconfinement phase transition.
 Our results are shown in Fig.~1b where we have taken the $F$-values at
 $R_0=0.55fm$ rather than at infinity due to the finite physical size of the lattices used in our
 simulation. The order parameter  nearly vanishes below $T_c$ and increases rapidly above $T_c$ indicating the
 phase transition. We expect $L^{ren}$ to have
 a well defined continuum limit because the short distance normalization to the
 $T=0$ potential is independent of the lattice cut-off [1]. In
 fact, through this short distance matching we have defined a renormalized
 Polyakov loop $L^{ren}$.
\vspace{-0.8cm}
\begin{figure}[ht]
\begin{center}
   \epsfig{file=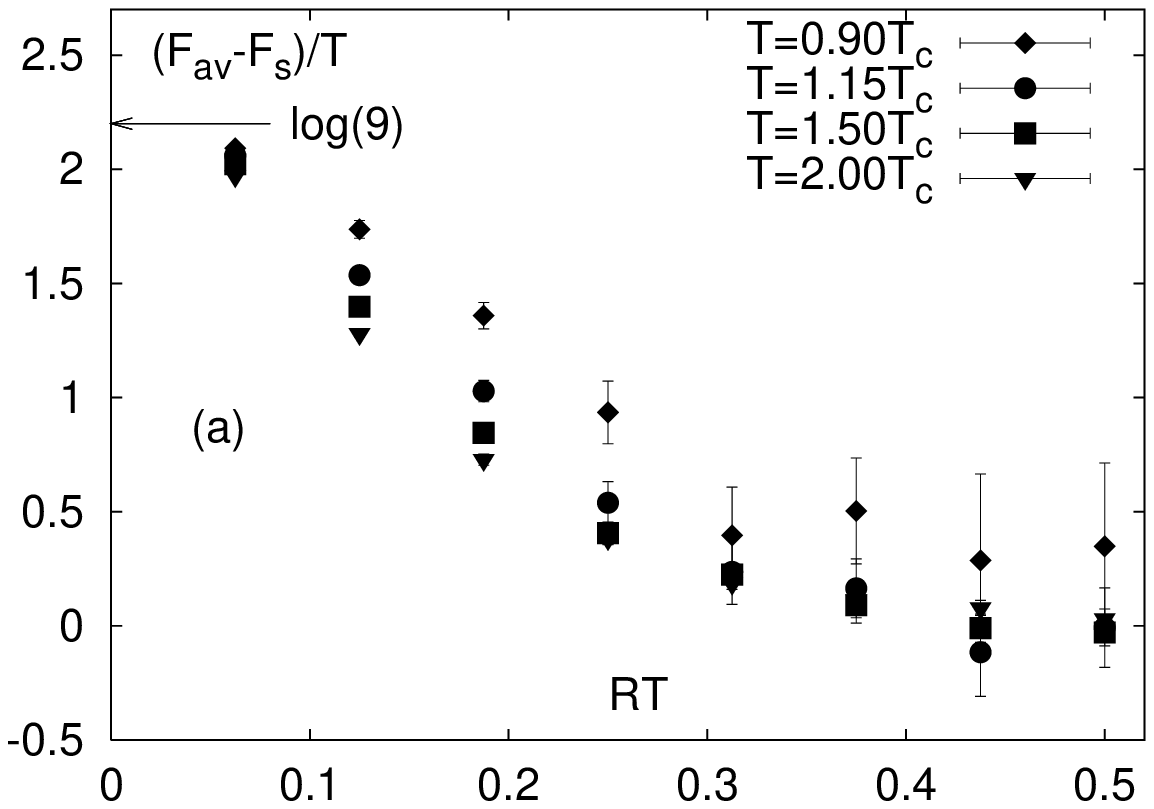,width=50mm}
   \hspace{2cm}
   \epsfig{file=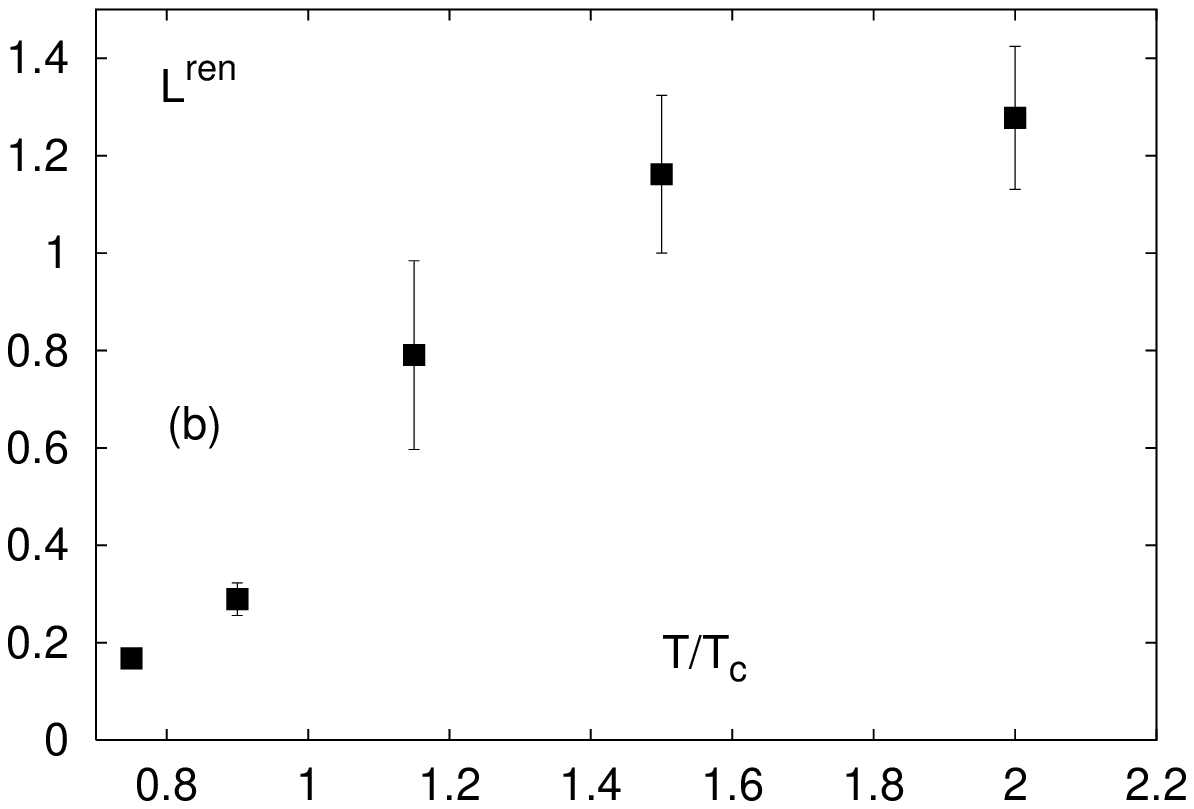,width=50mm}
\end{center}
\vspace{-1.6cm}
\caption{The relation between the color singlet and color averaged free energy
   as a function of $RT$ (a) and the new order parameter as a function of
   $T/T_c$ (b).}
\vspace{-0.7cm}
\end{figure}
\vspace{-0.5cm}
\section{CONCLUSIONS}
\label{section:results}
\vspace{-0.2cm}
We have shown that color averaging influences the free energy in both phases. Thus there
exists a non-vanishing octet contribution to $F_{av}$ in
the confinement phase. Our data indicate that the entropy contributions to the
free energy is $R$ and $T$ dependent, which makes a straight forward extraction
of the potential from free energies difficult.
We have constructed an order parameter for the deconfinement phase transition which has a well
defined continuum limit. It is an interesting question how the
discontinuity of the first order $SU(3)$ transition becomes visible in this
order parameter. 

\vspace{-0.2cm}


\begin{thebibliography}{99}
\vspace{-0.3cm}
\bibitem{} F. Zantow et al, hep-lat/0110103

\bibitem{}  L. G. McLarren, B. Svetitsky, \PR {\bf D24} (1981) 450

\bibitem{} S. Nadkarni, \PR {\bf D33} (1986) 3738

\bibitem{} L. S. Brown, W. I. Weisberger, \PR {\bf D20} (1979) 3239

\end{thebibliography}
\end{document}